\documentclass{aa}
\usepackage{graphicx}
\usepackage{txfonts}

\begin{document}
   \title{The nature of the recent extreme outburst of the Herbig Be/FU Ori binary Z\,CMa\thanks{
                 Based on observations made with ESO Telescopes at the 
                 Paranal Observatories under programme IDs: 
                 074.C-0442, 081.C-0410,  and 282.C-5041.}
}

   \author{
     T. Szeifert\inst{1}, S. Hubrig\inst{2}, M. Sch\"oller\inst{3}, 
     O. Sch\"utz\inst{1}, B.~Stelzer\inst{4}, 
     Z.~Mikul\'{a}\v{s}ek\inst {5}
   }
  
   \offprints{Thomas Szeifert}
   
   \institute{European Southern Observatory, Alonso de Cordova 3107, Casilla 
     19001, Santiago 19, Chile\\
     \email{tszeifer@eso.org}
     \and
     Astronomisches Institut Potsdam, An der Sternwarte 16, 14482 Potsdam, 
     Germany
     \and       
     European Southern Observatory, Karl-Schwarzschild-Str.\ 2, 85748 Garching,
     Germany
     \and
     INAF-Osservatorio Astronomico di Palermo, Piazza del Parlamento 1, 90134 
     Palermo, Italy
     \and       
     Department of Theoretical Physics and Astrophysics, Masaryk University, 
     Brno, Czech Republic
   }        

\authorrunning{Szeifert et al.}

   \date{\today --- the date to be modified later}
   \abstract
    {Z\,CMa is a binary system which consists of two young stars: A 
      Herbig AeBe component ``Z\,CMa\,NW'' embedded in a dust cocoon and a
      less massive component ``Z\,CMa\,SE'', which  is classified
      as a FU Orionis type star. Associated to the binary system is a giant 
      parsec size jet. Past spectropolarimetric observations showed that the 
      position angle of the linear optical polarization is perpendicular to 
      the jet axis, indicating that the visual 
      light escapes the cocoon via cavities aligned with the jet axis and is
      then scattered back into the line of sight of the observer.
      Recently, the system showed the largest outburst reported
      during the almost 90 years of available observations.
    }
    {We present new spectrophotometric and spectropolarimetric data obtained 
      in 2008 during the recent outburst phase.}
    {The data obtained in the visual spectral range at medium spectral 
      resolution were used to study the geometry of the system from the linear 
      polarization spectra and its magnetic field from the circular 
      polarization spectra.}
    {During the recent outburst we detect that the Z\,CMa system is polarized 
      by 2.6\% in the continuum and emission line spectrum, with a 
      position angle still perpendicular to the jet. From the high level of 
      polarization we 
      conclude that the outburst is associated with the dust embedded Herbig 
      AeBe NW component. The deep absorption components of the Balmer lines in 
      the velocity frame extending from zero velocity reaching a wind velocity 
      of $\sim$700\,km\,s$^{-1}$, together with the absence of a red-shifted 
      broad emission at similar velocities, indicate a bi-polar wind.
      We do not detect a significant mean longitudinal magnetic field during 
      the outburst, {\bf but the data obtained in 2004 indicates the possible 
      presence of a rather strong magnetic field of the order of $\sim$1\,kG.}
      However, we critically review the applied method of magnetic field 
      measurements in the presence of a strong stellar wind. 
      {\bf The main result of our studies is that the bolometric luminosity of 
      Z\,CMa remained surprisingly constant during the recent ``outburst''.
      We conclude that either the geometry of the cavity through which the 
      light escapes from the cocoon has opened a new path, or that the screen 
      of dust, which reflects the light toward the observer became more 
      efficient causing the observed increase of the visual brightness by 
      about 2\fm{}5.}
    }
    {}
   \maketitle

\section{Introduction}

Z\,CMa (HD 51370) is a young visual binary system consisting of a 
12\,M$_\odot$ Herbig AeBe star and a 3\,M$_\odot$ FU\,Ori companion at an age 
of 0.3\,Myr (e.g.\ Alonso-Albi et al.\ \cite{Alonso2009}). Both components 
are surrounded by active accreting disks. Associated with the binary system is 
a giant parsec size jet (Poetzel et al.\ \cite{Poetzel1989}). First evidence 
for the binarity was an elongation of the K-band image of this star toward the 
northwest (Leinert \& Haas \cite{Leinert1987} and  
Koresko et al.\ \cite{Koresko1989}). The binary was resolved by 
Koresko et al.\ (\cite{Koresko1991}) with near-IR speckle interferometry. 
The angular separation of the binary was 
only $\sim$0\farcs1 with a position angle of 120$^{\circ}$. At that time the 
south-eastern (Z\,CMa\,SE) component was dominant in the visual spectral range 
and near-infrared
$J$ and $H$ bands, while the north-western component was found brighter at 
2.2\,$\mu$m and longer wavelengths. Thi\'ebaut et al.\ (\cite{Thiebaut1995}) 
resolved the binary system with speckle interferometry at 730\,nm and 656\,nm. 
From the data obtained in the quiescent state in 1989.84 the authors deduced 
$L$ = 1300\,$L_{\odot}$, $R$ = 13\,$R_{\odot}$ and 
$T_{\rm eff}$=10000\,K for the Z\,CMa\,SE FU\,Ori component
and $L$ = 2400\,$L_{\odot}$, $R$ = 1690\,$R_{\odot}$, and $T_{\rm BB} = 980$\,K
for the near infrared Z\,CMa\,NW component.
The authors attributed the excess flux associated with Z\,CMa\,NW at 
optical wavelengths to scattered light, arguing that photons originating from 
the hidden central source are scattered back into the line of sight. 
Scattering of the photons is assumed to take place along the walls of a 
cavity (usually ascribed to a strong bipolar outflow) evacuated within the 
envelope. Radio observations in 2005 and 2006 in the cm and mm range suggest 
that we see Z\,CMa\,NW at an inclination of 30$^{\circ}$ (Alonso-Albi et al.\ 
\cite{Alonso2009}). This means that we do not observe the inner walls of the 
bipolar cavity within the dusty envelope, where the photons are scattered. 

The scattered light scenario is well consistent with optical 
spectropolarimetric observations of Whitney et al.\  (\cite{Whitney1993}), 
who detected in 1991--1992 a significant linear polarization of 1\% to 2\% in 
the continuum and up to 6\% in the strong emission lines. The different
polarization of continuum and lines have led Whitney et al.\ 
(\cite{Whitney1993}) to suggest that the emission lines have their origin in 
the Z\,CMa\,NW component. At that time the NW component contributed about 20\% 
of the flux of the whole system at visible wavelength. The Z\,CMa\,SE 
component is assumed to be unpolarized, the NW component is assumed to
show 6\% linear polarization and the composite spectrum of the binary would
then depend on the flux ratio at the respective wavelength. The position 
angle of the polarization of 150$^{\circ}$ in November 1991 and 154$^{\circ}$ 
in March 1992 was roughly perpendicular to the 3.6 parsec scale jet of 
Z\,CMa at a position angle of 60$^{\circ}$, discussed by 
Poetzel et al.\ (\cite{Poetzel1989}).
The same authors reported several velocity components in the jet of which 
the fastest knots reaches a velocity of about $-620$km\,s$^{-1}$.
This position angle of polarization indicates that the
light from the Z\,CMa\,NW component primarily escapes from the envelope
by scattering off the walls of a jet-blown cavity. A smaller scale jet with a 
similar position angle was recently discovered in X-rays by Stelzer et al.\ 
(\cite{Stelzer2009}).

\begin{figure}[tp]
\centering
\includegraphics[width=8.5cm,angle=0.0,clip=,bbllx=30pt,bblly=60pt,bburx=560pt,bbury=380pt]{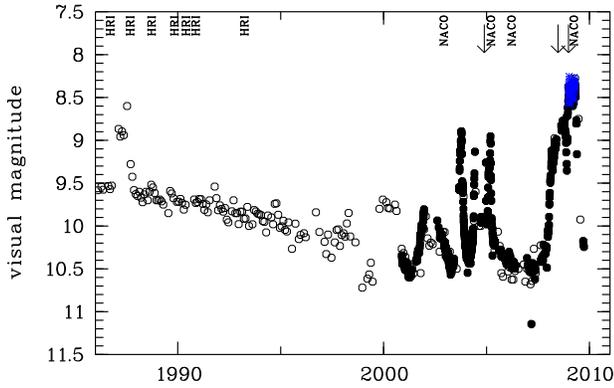}
\caption[]{Light curve of Z\,CMa over the last decades. Open symbols denote 
the observations from American Associations of Variable Stars Observers 
(AAVSO). 
The filled symbols correspond largely to the ASAS data and during the
peak of the lightcurve {\bf to our measurements (blue asterisks)}.
Epochs of NACO observations, the high resolution images as summarized by 
Thi\'ebaut et al.\ (\cite{Thiebaut1995}) and the epoch of spectropolarimetric 
observations (arrows) are indicated on the top.}
\label{lightcurveshort}
\end{figure}

According to photometric and visual observations available from various 
archives, Z\,CMa exhibited recently the largest outburst reported 
during the almost 90 years of available observations.
The light-curves presented in Fig.~\ref{lightcurveshort} were extracted from 
the archives of the ``All Sky Automated Survey'' (ASAS) rejecting the data 
points with quality flags C and D (Pojmanski\ \cite{Pojmanski2002}).  
More data over a larger time span were retrieved from the AAVSO data base. 
For the later visual estimates the data points taken for a given month 
were averaged after clipping data points beyond a 2$\sigma$ threshold.
Irregular variability was observed in the thirties of the last century. Later 
Z\,CMa steadily increased its brightness, reaching a phase of constant 
brightness until about 1973. Since then the star was fading from almost 
9$^m$ in 1973 to 10\fm5 in 2007.  Superimposed to the fading lightcurve we can 
identify in Fig.~\ref{lightcurveshort} several more recent outbursts, the 
most prominent ones were observed in 1987 and 2000 with a duration similar to 
that of eruptions of EX Lupi-type (EXors) young stars. Between 
2002 and 2006 the light-curve appeared very irregular. 
A gradual increase of the visual brightness started at the end of 2007, 
reaching the highest brightness between November 2008 and May 2009. This 
bright phase has now ended, with the latest observations reporting a visual 
brightness fainter than 10$^m$ after a very rapid decline.

In this letter we present new optical spectropolarimetric observations of 
linear and circular polarization in the Z\,CMa system, obtained during the 
recent outburst, and discuss the cause for this outburst.

\begin{figure}
\centering
\includegraphics[width=8.4cm,angle=0.0,clip=,bbllx=90pt,bblly=45pt,bburx=555pt,bbury=350pt]{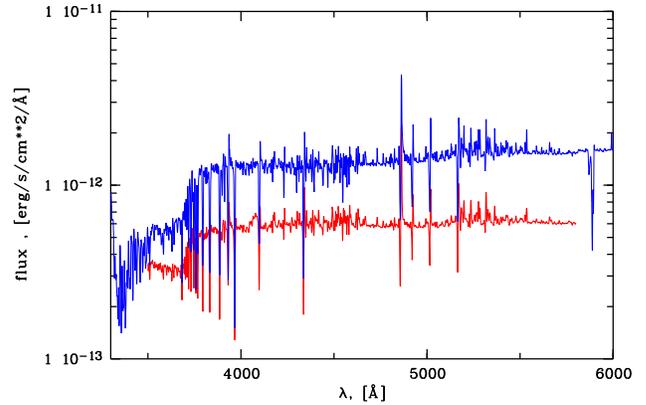}
\caption[]{Spectrophotometric flux of Z\,CMa in 2004 (lower spectrum) and 
2008 (upper spectrum).}
\label{specphot}
\end{figure}

\begin{figure}
\centering
\includegraphics[width=6.6cm,angle=0.0,clip=,bbllx=70pt,bblly=30pt,bburx=570pt,bbury=400pt]{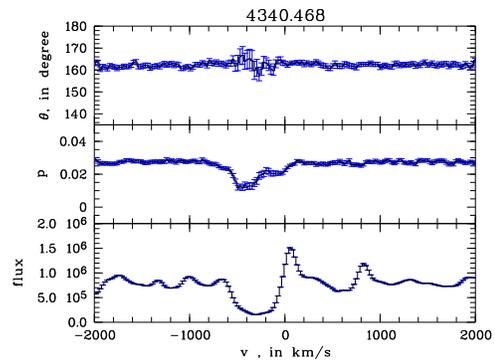}
\caption[]{Spectropolarimetry of Z\,CMa in H$_{\gamma}$.}
\label{spolplot4340}
\end{figure}

\begin{figure}
\centering
\includegraphics[width=6.0cm,angle=0.0,clip=]{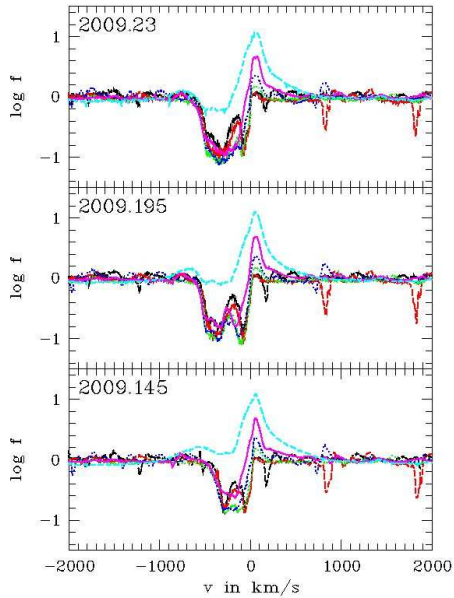}
\caption[]{Logarithmic spectral shape of the Balmer lines H$_{\alpha}$ to
 H$_{9}$. The variability is most pronounced in the absorption troughs
and for H$_{\alpha}$ in the symmetric electron scattered broad wings.}
\label{spec6562}
\end{figure}

\section{Results and Discussion}

\subsection{Spectroscopy and Linear Polarization}

The spectropolarimetric data sets provide very important insights into the 
nature of the at visual wavelengths unresolved binary star. We found 
significant changes of the line shapes in the linearly polarized light with 
respect to the data published by Whitney et al.\ (\cite{Whitney1993}). In our 
data obtained during the extreme outburst, we see a flat polarization spectrum 
with an average polarization P=2.6\% $\pm$0.1\% at 4350\AA\ at a position 
angle of 160$^{\circ}\pm1^{\circ}$.
This polarization angle is roughly perpendicular to the parsec scale jet 
associated with Z\,CMa, and similar to observations presented 
by Whitney et al. (\cite{Whitney1993}) who measured in November 1991
P = 2.0\%, P.A = 150$^{\circ} \pm 1.5^{\circ}$  and four months later  
P = 1.4\%, P.A = 154$^{\circ} \pm 1.5^{\circ}$. 
However, in the deep absorption troughs 
of the Balmer lines and other strong wind absorption lines, we detect 
depolarization signatures (see Fig.~\ref{spolplot4340}). These line 
components caused by the accelerated wind show a composite spectrum: 
We observe a  broad blue trough typically associated with an accelerated
wind, visible to about $-$600\,km\,s$^{-1}$, with less deep absorption up to 
about $-$800\,km\,s$^{-1}$. As is shown in more detail in Fig.~\ref{spec6562}, 
this is not mirrored by emission on the red side of the lines as would be 
expected for spherical symmetric winds. The trough is rather deep, indicating 
that a large fraction of the stellar disk is obscured by the wind acceleration 
zone. 
Superimposed to the wind feature is a very narrow emission component of 50 to 
60\,km\,s$^{-1}$ half width, which can be seen as well for faint low 
excitation lines like of Fe\,{\sc ii} and similar species. In contrast to the 
observations of Whitney et al.\ (\cite{Whitney1993}), we find that the narrow 
emission lines show the same degree (2.6\%) and position angle of polarization 
as the continuum. These observations can be understood within the above 
mentioned model that the visible light can only escape via cavities from the 
dust cocoon, most likely aligned with the bi-polar jet and a changing
flux ratio between the two stars:
In our recent observations the continuum radiation as well as 
the emission components are dominated by the NW component, while only in the 
deep absorption troughs the unpolarized SE component is still contributing. 
Whitney et al.\ (\cite{Whitney1993}) observed the binary in a different state, 
when the unpolarized SE component dominated the flux in the near-UV and 
visual. The strongest emission lines primarily emitted by the NW-component 
reached up to a polarization of about 6\%. This high value must be seen as 
the degree of polarization of the NW-component at all wavelength and 
accordingly the polarization of the NW-component was higher at that time than 
what we see now in the visual and near-UV continuum.
The change of the degree of polarization of the NW component from  6\% at
a low state to 2.6\% at the brightest state, together with the brightening of 
the target indicates that the geometry of the cavity through which the light 
escapes from the cocoon has opened a new path, or that the screen of dust, 
which reflects the light toward the observer, became more efficient, 
causing an increase in the visual brightness by about 2\fm5.
In the strongest Balmer lines the absorption trough presents a double
plateau with the deepest depolarization starting at the terminal 
wind velocity of about $-$600\,km\,s$^{-1}$ and ending at about 
$-$200\,km\,s$^{-1}$. At red-shifted wavelengths the NW component 
dominates the composite spectrum. It is possible to explain the double plateau 
on the blue-shifted side, if the spectrum of the SE component is that of
a fast rotator with hydrogen absorption lines such that the contribution of 
the SE component to the composite spectrum is smaller from $-$200 to 
0\,km\,s$^{-1}$ at wavelengths. 
Other line species which show depolarization signals in the blue-shifted
parts of line profiles are calcium lines in which the depolarization
signal can only be seen at $v < -200$\,km\,s$^{-1}$ and the Fe\,{\sc ii}-lines,
which do not present the deep absorption troughs in the flux spectrum,
but show a double absorption feature in the linear polarization
with an absorption at the velocity of the fast wind and another absorption
at the system velocity. For the later we can not conclude if the 
depolarization has an intrinsic origin in the NW component, or if
the presumed young FU\,Ori star contributes
unpolarized line emission to the composite spectrum.

We note that the depth and the width of the absorption at the blue-shifted
side of the Balmer lines, in absence of a broad 600\,km\,s$^{-1}$ red-shifted 
emission, indicates a bipolar geometry of the wind. If no such broad red 
emission is observed, then it can be argued that either there is no gas 
streaming out except out of the polar region, or that the recombination 
radiation in the line is not visible through the cavities, but blocked by the 
disk or cocoon. For the H$\alpha$ line we find however a broad emission 
component beyond the velocity of the wind absorption feature. The 
H$\alpha$ absorption trough is not as deep as for other Balmer 
lines. This broad component is caused by electron 
scattering as is often seen in luminous hot stars and therefore not 
related to high velocity outflows.

The spectral slope as indicated in Fig.~\ref{specphot} appears
redder during the recent outburst, which contradicts an increase of 
temperature as source of the luminosity change as
proposed by van den Ancker et al.\ (\cite{vandenAncker2004}). Similarly, the 
very insignificant change of the depth of the Balmer jump would argue
against a dramatic change of stellar parameters.

\begin{figure}
\centering
\includegraphics[width=8.4cm,angle=0.0,clip=]{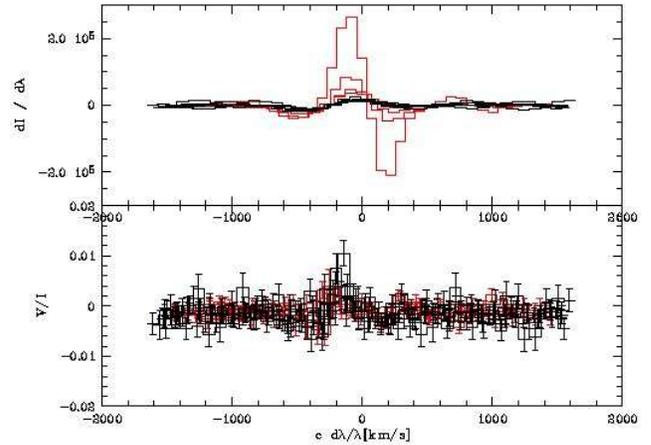}
\caption[]{Circular polarization in the Balmer line spectra observed in 
November 2004 (on the bottom) together with the spectral gradients 
d$I /$d$\lambda$.
The spectra are plotted in the velocity frame to display Balmer lines from 
4861\,\AA{} to 3734.370\,\AA{}. Both curves indicate that the s-shaped 
profiles of the high-n Balmer lines are not centered at the rest frame 
wavelengths. For the low-n Balmer lines, in this figure strongly deviating in 
d$I/$d$\lambda$ from the high-n profiles, the peak of the emission is 
close to the rest frame wavelength. Accordingly the spectral gradient is
zero at the rest frame wavelength and doesn't correlate with the Stokes\,V
profile.}
\label{circ53331}
\end{figure}

\onlfig{6}{
\begin{figure}
\centering
\includegraphics[width=8.4cm,angle=0.0,clip=]{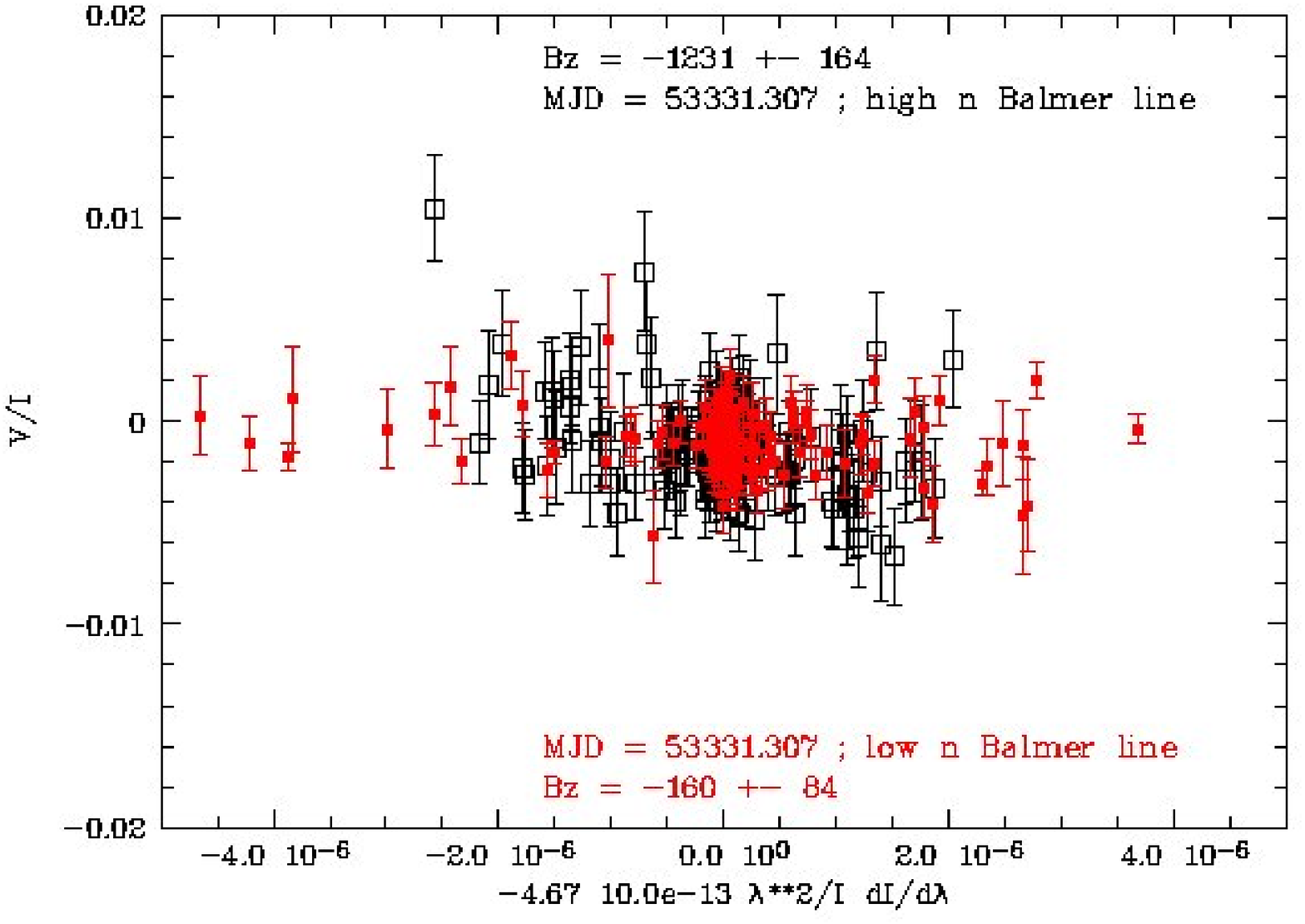}
\caption[]{Magnetic field measurements using the data obtained in November 2004.
The longitudinal magnetic field is derived following the equation
$V/I = -{\rm g}_{\rm eff} {\rm C}_z \lambda^2 \frac{1}{I} \frac{{\rm d} I}{\rm{d} \lambda} \langle B_z \rangle$ from
the slope of the linear fit through the data points (see 
Bagnulo et al.\ \cite{Bagnulo2002}). Large open 
and small filled symbols are obtained from the high number and low number 
Balmer lines, respectively.}
\label{field53331}
\end{figure}
}

\subsection{Circular Polarization}

The analysis of circular polarization measurements obtained during the 
outburst reveals no significant detection. In the data obtained in May 2008 
we detect a weak signal of circular polarization in the trough of all Balmer 
lines, which however does not match in wavelength the unpolarized line profile.
However, while reviewing the archive data already published by Wade et al.\ 
(\cite{Wade2007}) we do detect a clear circular polarization signal in 
particular at the blue shifted side of the high-n Balmer lines. For the 
low excitation lines from H$\beta$ to H$\epsilon$ we measure $-160 \pm 84$\,G, 
while from the higher excitation Balmer lines a very significant field of 
$-1231 \pm 164$\,G (see Figs.~\ref{circ53331} and \ref{field53331}) was 
obtained. Still, these results have to be taken with some caution.
To properly interpret the circular polarization signal, it is necessary to
consider the composite nature of the binary system spectrum, 
the superimposed effects of the stellar wind, and the variability of the
target. The B-field diagnostics as applied above correlates the s-shaped
profile in  $V/I$ with the s-shaped profile of the spectral gradient
d$I/$d$\lambda$ of the Voigt profiles.
The B-field diagnostics may, for instance, be disturbed by the
light of the companions, and, in particular, by the stellar wind features due 
to the spectral gradient d$I/$d$\lambda$ along the x-axis of the diagnostic 
diagram (see Figs. ~\ref{circ53331} and ~\ref{field53331}).
It should as well be considered that the target is variable on very short time 
scales, such that spectra may have changed between the respective exposures at 
$\pm45^{\circ}$.

It is as well unclear if these fields were detected in the dust
embedded Herbig Ae star or in the SE companion. In the first case
the strong field is only seen in the fainter Balmer lines, because
the high opacity of the expanding wind at the blue shifted part 
of the strong lines would not allow light from the photosphere
to escape from the NW component. Alternatively, without any indication 
of the contribution of the respective two stars to the UV and 
visible spectra, one can as well explain the data if the 
magnetic field was detected in the SE component. This component may have 
dominated the spectra of 2004 in the wavelength range of the Balmer
lines where we measure the Stokes V features, whereas later in May 2008, 
the contribution of Z\,CMa\,SE became insignificant on the blue-shifted 
side of the P\,Cygni line of Z\,CMa\,NW such that we detect a hint
of circular polarization in the red-shifted part of the hydrogen lines only,
but no S-shaped feature, which would lead to a magnetic field detection.
Nor would the wind dominated spectral line gradients d$I/$d$\lambda$ linearly 
correlate with $V/I$. With the few available spectra it is not yet possible 
to reliably conclude on the origin of the features in the circular 
polarization spectrum.

\subsection{Origin of the Outbursts}

Despite of the different origin of the NIR Ks band flux from the 
surrounding dust cocoon and the optical flux with an origin from 
the pole of the star, we presume that a strongly changed optical/UV 
source would heat up the dust and the increase of the luminosity of 
the stellar core would be re-emitted by thermal radiation in the 
Ks-band and beyond. There is however no significant change of the 
Ks-band flux as far as we can tell from the K-band flux ratios given
in Table~\ref{orbits:2} or from comparing the most recent report of infrared
observations by Antoniucci et al.\ (\cite{Antoniucci2009}) of Ks$=3.56$
at the peak of the lightcurve in March 2009,
with Ks$=3.79$ reported by Koresko et al.\ (\cite{Koresko1991}).
Similarly there is no change in the X-ray properties reported by 
Stelzer et al.\ (\cite{Stelzer2009}) in respect to earlier measurement
at a `quiescent' phase. If the short-term optical brightening was an 
EXor-like accretion event, X-ray variability would have been expected. 
Our polarimetric data leads to a new interpretation in which 
the X-ray emission associated with the binary may be attributed to the 
lower-mass SE component which is not involved in the outburst.

The apparently insignificant change of the K-band thermal emission
rather argues against the strongly increased luminosity of 
 $3\times10^{5}$\,$L_{\odot}$ during short outbursts, as estimated by van den
Ancker et al.\ (\cite{vandenAncker2004}) from the visual SEDs with respect 
to only  $L = 2400$\,$L_{\odot}$ estimated by Thi\'ebaut et al.\ 
(\cite{Thiebaut1995}) from the SED including the thermal emission.
It rather underlines our interpretation that the bolometric luminosity
during the outbursts remains constant, but the amount of visible light
transmitted through the envelope has changed. 
The derived high luminosity was primarily a consequence of the
early spectral type estimated for the NW source from the He\,{\sc i} and
O\,{\sc i} lines, which lead to very high bolometric and extinction 
corrections. We detect these lines with blue shifted broad absorption profiles 
and with P\,Cygni profiles respectively, both indicating an origin of the 
lines in the accelerated wind rather than in the photosphere of the NW 
component. The wind however, with indication of bipolarity and eventually with 
indications for a magnetic field, may well be excited by other 
sources rather than by the thermal radiation from the star.
The short-term outbursts are nevertheless superimposed to long-term
FU\,Ori like changes in the visual brightness, with a slow brightening by 
one magnitude until about 1970 and a subsequence fading by a similar amount. 
It is well possible that any of the binary components contributes to these
changes.

\section{Summary}
We have shown that the NW component of Z\,CMa was undergoing a historically
bright outburst in 2008/2009. Our polarimetric data supports a variable 
scattering geometry of the NW envelope as the cause for the optical outburst, 
rather than FU\,Ori or EXor events from the SE component.
From the changes in the linear polarization spectra 
taken at the outburst phase we conclude that the scattering path in which the 
light of Z\,CMa\,NW escapes from the dust cocoon must have changed. The
linear polarization from the now dominant NW component was found smaller
than the 6\% estimated at the fainter phase in 1991.
We do not find changes in the K-band magnitude in contrast to the recent 
change of 2\fm5 in the visual. A significant fraction of the variability 
may be explained by the changed, more direct line of sight in which we now 
see the Z\,CMa\,NW:
More light from the central source would reach the observer through a
widened cavity, while the central source directly observed in the K-band 
did not significantly change in luminosity.

\begin{acknowledgements}
We acknowledge with thanks the variable star observations from the AAVSO 
International Database contributed by observers worldwide and used in this 
research. This work profited greatly from data extracted from the ASAS
data base. 
Based on observations made with the European Southern Observatory
telescopes obtained from the ESO/ST-ECF Science Archive Facility.
\end{acknowledgements}

\Online

\begin{appendix}

\begin{table}
\caption[]{Journal of spectroscopic observations.}
\label{spectroscopy:1}
\begin{tabular}{cccl} \hline\hline
\multicolumn{1}{c}{Date} &
\multicolumn{1}{c}{Instrument} &
\multicolumn{1}{c}{$\lambda$ range [\AA]} &
\multicolumn{1}{c}{Detector, grism} \\ \hline\noalign{\smallskip}
2004/11/22 & FORS\,1  & 3470 - 5870 & TEK, 600B\\
2008/05/22 & FORS\,1  & 3310 - 6200 & E2V, 600B\\
2008/12/26 & FORS\,1  & 3310 - 6200 & E2V, 600B\\
2008/12/26 & FORS\,1  & 3680 - 5120 & E2V, 1200B\\
2009/02/23 & FEROS  & 3570 - 9210 & Echelle\\
2009/03/13 & FEROS  & 3570 - 9210 & Echelle\\
2009/03/26 & FEROS  & 3570 - 9210 & Echelle\\
\hline
\end{tabular}
\end{table}

\begin{table}
\caption {Journal of NACO observations. The position angle
is given with respect to the SE component. For the L-band data the
angular separation and position angle was set to a fixed value to obtain
a reliable flux ratio.
}
\begin{tabular}{cccccl}\hline\hline
  Year   & Band    &  f$_{NW}/$f$_{SE}$  &    PA [$^{\circ}$]     &    d [\arcsec{}]       \\ \hline\noalign{\smallskip}
2002/12/05 & L   & 49.57 $\pm$0.15&        &         \\
2002/12/05 & J   &  0.94 $\pm$0.01& 309.90 &  0.107  \\
2002/12/05 & Ks  &  3.97 $\pm$0.01& 311.9  &  0.102  \\ 
2005/02/16 & Ks  &  4.7  $\pm$0.01& 311.43 &  0.098  \\
2005/02/16 & H   &  2.79 $\pm$0.01& 310.98 &  0.114  \\
2005/02/16 & NB2.17&4.34 $\pm$0.05& 314.74 &  0.104  \\
2006/03/13 & NB2.12&3.63 $\pm$0.01& 311.93 &  0.101  \\
2006/03/13 & NB2.17&3.42 $\pm$0.01& 312.43 &  0.101  \\
\hline
\end{tabular}
\label{orbits:2}
\end{table}

\section{Observations and Data Reduction}

The observational material discussed in this section includes FORS\,1 
spectropolarimetric data (Appenzeller et al.\ \cite{Appenzeller1998}) obtained 
in the framework of our programs carried out in May 2008 and in December 2008. 
Additionally, we used ESO archive data from 2004 obtained in circular 
spectropolarimetric mode. These data were used by Wade et al.\ 
(\cite{Wade2007}) to search for the presence of longitudinal magnetic fields 
in a larger number of Herbig AeBe stars. These older observations (see 
Table~\ref{spectroscopy:1}) were carried out with a slit width of 0\farcs{}8 
and grism 600B ($R\approx1000$). The flux spectra in the ordinary and 
extraordinary ray of the Wollaston prism were extracted for all position 
angle settings of the retarder waveplate. Since there were no available flux 
calibrators observed for this setup in the respective night, we used flux 
calibrators taken in normal spectroscopic mode without the Wollaston prism. 
The new spectropolarimetric data from 2008 were obtained with the same 
instrument, but now equipped with a higher UV response detector system with a
smaller pixel size.  During the December run we have used the higher dispersion
grism 1200B in addition to the data with the 600B grism. In all our 
observations the slit width was set to 0\farcs{}4 to optimize the resolution 
for the polarimetric measurements ($R\approx4000$ and $R\approx2000$, 
respectively).
The spectrophotometric calibrators retrieved from the archive were observed 
without polarization optics on different nights. One source of uncertainty 
in the flux calibration is therefore the unclear spectral response of the 
wave retarder plates, for which we presume a high throughput over the full 
spectral range. However, significant slit losses are expected with this 
setup due to the narrow slit widths and the difficult acquisition of the 
target on the slit considering the extreme brightness of the target for 
8\,m-class telescopes. To remedy this difficulty related to the absolute 
calibration, we re-scaled the flux at 5500\,\AA{} to the monthly average of the 
AAVSO visual observations of 9\fm4 in November 2004 and 8.4 in 
December 2008. 
In Fig. \ref{specphot} we present spectrophotometric 
fluxes for the Z\,CMa system in 2004 and 2008, respectively.
Further, for the proper interpretation of the measured polarization angle, we
corrected the result with the chromatic zero angle of the
super-achromatic half wave retarder plate.\footnote{The data is available 
on the instrument web pages: 
http://www.eso.org/sci/facilities/paranal/instruments/fors/ .} 
To perform linear polarization measurements, a Wollaston prism and a half-wave 
retarder plate rotated in 22.5$^\circ$ steps between 0 and 67.5$^\circ$ were 
used. For circular polarization measurements the quarter-wave retarder plate 
was used at the positions $+$45$^\circ$ and $-$45$^\circ$. 
Our FORS\,1 data were supplemented by three high-resolution 
FEROS spectra ($R$=48\,000) obtained in February and March 2009 to study the 
behavior of hydrogen lines. The journal of spectroscopic observations is 
presented in Table~\ref{spectroscopy:1}.

To better understand the parameters of the resolved binary system we retrieved 
all available  NACO adaptive optics near-infrared data (Lenzen et al.\ \cite{Lenzen2003}) from the ESO archive.
As for the FORS\,1 data sets, the target is rather bright for imaging mode 
observations with 8\,m-class telescopes and accordingly the various data sets
were taken with a mixed choice of neutral density and narrow band filters.
As a consequence we cannot provide absolute photometry for NACO
data with the required level of confidence, but only the flux ratio of
the two components (see Table~\ref{orbits:2}) and the astrometric
parameters of the binary. 
In case of normal jittered observation sequences we have used the
Eclipse data reduction package provided by ESO to NACO users. In other
cases we had only two acquisition exposures or a few jitter positions.

\end{appendix}
\end{document}